\documentclass[a4paper]{article}

\usepackage{INTERSPEECH2016}

\usepackage{graphicx}
\usepackage{amssymb,amsmath,bm}
\usepackage{textcomp}
\usepackage{lipsum}

\usepackage{multirow}
\usepackage{url}
\usepackage{slashbox}
\usepackage{dirtytalk}

\ninept

\title{Realistic multi-microphone data simulation \\
for distant speech recognition}

\makeatletter
\def\name#1{\gdef\@name{#1\\}}
\makeatother \name{{\em Mirco Ravanelli, Piergiorgio Svaizer, Maurizio Omologo}}

\address{Fondazione Bruno Kessler (FBK), Trento, ITALY \\
  {\small \tt mravanelli@fbk.eu, svaizer@fbk.eu, omologo@fbk.eu}
}

\begin{document}

  \maketitle
  \begin{abstract}
The availability of realistic simulated corpora is of key importance for the future progress of distant speech recognition technology. The reliability, flexibility and low computational cost of a data simulation process may ultimately allow researchers to train, tune and test different techniques in a variety of acoustic scenarios, avoiding the laborious effort of directly recording real data from the targeted environment.

In the last decade, several simulated corpora have been released to the research community, including the data-sets distributed in the context of projects and international challenges, such as CHiME and REVERB.
These efforts were extremely useful to derive baselines and common evaluation frameworks for comparison purposes. At the same time, in many cases they highlighted the need of a better coherence between real and simulated conditions.

In this paper, we examine this issue and we describe our approach to the generation of realistic corpora in a domestic context. Experimental validation, conducted in a multi-microphone scenario, shows that a comparable performance trend can be observed with both real and simulated data across different recognition frameworks, acoustic models, as well as multi-microphone processing techniques.
  \end{abstract}

 \section{Introduction} \label{sec:intro}
Distant Speech Recognition (DSR) represents a fundamental technology towards natural human-machine interfaces.  Despite the recent  substantial progress in various related fields, including spatial filtering \cite{BrandWard,beam}, microphone selection \cite{nadeu}, source separation \cite{bss}, speech dereverberation \cite{derev}, speaker localization ~\cite{DeMori98}, acoustic event detection \cite{aed1} as well as acoustic modeling \cite{pawel2,hain,dnn_rev,dnn_rev2,dnn3}, DSR still exhibits a  lack of robustness, especially when adverse acoustic conditions originated by non-stationary noises and acoustic reverberation are met \cite{adverse}.  

The availability of realistic simulated corpora and, more importantly, the definition of  common methodologies, algorithms and good practices  to generate simulated data plays a crucial role for fostering future research in this field and will eventually help  researchers to better migrate laboratory results into real application scenarios.  
%MO 24/6
%The flexibility offered by data simulation can indeed help researchers in evaluating novel techniques in different scenarios, possibly considering different acoustic environments, microphone configurations as well as acoustic conditions, without the need of expensive and time consuming recordings in the targeted environment. Moreover, considering the low computational cost required by the simulation process, such approach is also very helpful for training purposes.  
Approaches as contaminated speech training \cite{ravanelli15,matassoni,rav_in14}, multi-style training \cite{mstyke,cont2,cont3} and data augmentation \cite{dataaug,nakatani,dataaug2,dataaug3} have, in fact, been shown very effective in improving the DSR system performance.

During the last decade, some simulated corpora have been made available to the research community under projects or international challenges. Valuable examples are the corpora released  under the ChiME \cite{chime,chime3} and REVERB \cite{revch} challenges, which have contributed to define common tasks, baselines and evaluation frameworks across researchers. Other simulated corpora have been released under the CHIL project \cite{chil_corpus} and, more recently, under the EU DIRHA project \cite{lrec,dirha_grid,hscma,dirha_icassp,dirha_asru}. 
These efforts were extremely important to stimulate research in the DSR field, but in several cases they also pointed out the need of a better coherence  between real and simulated data performance. In \cite{chime3}, for instance, the authors state that \say{\textit{The [CHiME3] challenge has drawn attention to the value of simulated training data, but highlighted the need for better simulation algorithm. It has also demonstrated that caution is needed when interpreting results of challenges that use simulated data evaluation.}}.
We fully agree with this statement, as our past experience confirms that prudence is needed when using simulated data. This caution is often to be attributed to very subtle differences that may characterize the process of simulation as, for instance, the accuracy and the realism of impulse responses. 

%It would be thus of great interest for the research community to establish common methodologies, algorithms and good practices for the generation of realistic simulated corpora, eventually helping  researchers to better migrate laboratory results into real application scenarios. 
%MO 24/6
%In this paper,  we face this issue and describe our methodology to the generation of realistic simulated corpora. An experimental validation, conducted in a multi-microphone domestic scenario and considering  different experimental conditions, has shown that the proposed approach leads to an interesting level of agreement in the performance of real and simulated data. %data   including different recognition frameworks, acoustic models, as well as multi-microphone processing.
The main purpose of this paper is to investigate on the level of agreement in performance trend, that can be obtained with real and simulated signals. A major focus of our work is on reverberation, rather than background noise.
Simulations are based on the contamination method described in \cite{matassoni}. Each impulse response (IR) is 
%24/6estimated 
measured according to the procedure described in \cite{Ravanelli-12}, while simulated IRs are derived by a modified version of the image method \cite{image}, which was experimented in our past works. This modified version differs from the original \cite{image} just for simulating also the directivity pattern of the source, besides sound propagation effects.
The experiments, conducted in a new multi-microphone domestic scenario that was developed under DIRHA, demonstrate a good level of agreement in performance, evident with all the investigated acoustic models and processing. We also show the improvement that can be obtained when 
%24/6 real 
measured IRs, instead of image-method based ones, are used to train acoustic models.

The paper is organized as follows: Sec. \ref{sec:datasim} outlines the data simulation approach; Sec. \ref{sec:expsetup} describes the adopted experimental setup, while Sec. \ref{sec:expres} reports on the experimental validation of the methodology. Finally, Sec. \ref{sec:conc} will draw some conclusions and provide an outlook on future work.

\section{Data Simulation} \label{sec:datasim}
In this work, the data simulation process is achieved according to the following equation\footnote{\url{https://github.com/mravanelli/pySpeechRev}}:

\begin{equation}
\label{eq:1}
y(t)=x(t)*h(t)+n(t)
\end{equation}
where $y(t)$ is the simulated distant-talking signal, $x(t)$ is the close-talking speech, $h(t)$ is the impulse response of the acoustic environment for a given source and microphone position, $*$ is the convolution operator, and $n(t)$ is an additive background noise. 
Several important aspects must be considered for an effective simulation, as discussed in the following.
%A detail discussion of the main aspects behind our data simulation approach is proposed in the following sections.
%Particular focus will be devoted on the main aspects we believe are of great importance to obtain realist simulated corpora.

\subsection{Close-Talking Recordings}  \label{sec:ctrev}
Our experience in data simulation suggests that the availability of a high-quality close-talking data set is crucial for generating realistic distant-talking simulated data. Particular attention should be directed to ensuring dry and noiseless recordings, since the possible presence of noise sources, saturation, reverberation effects due to the room acoustic as well as distance between speaker and microphone can produce artifacts in the later simulation process. The quality and the characteristics of the microphone can also influence the realism of the simulations.%For this reason, the role played by the recording environment as well as the quality of the adopted recording equipment are of crucial importance to obtain realistic simulation.

In the context of the DIRHA project, high quality close-talking speech signals have been acquired under extremely quiet conditions (with a SNR of at least 50-60 dB for each sentence), in an acoustically treated recording room, using a high-quality microphone (Neumann TLM 103) and a professional audio card (RME Octamic II). %with the final purpose of creating realistic simulated data in a domestic environment. %To enhance speech dryness, the recording studio was reinforced by absorption materials on the walls, floor and ceiling in order to limiting signal reflections. Moreover, particular
%Particular care was taken to avoid any noise and interference, both at acoustical and electrical level, ensuring a SNR of at least 40dB for each sentence.  
%The speakers were sitting in front of a display (showing the sentence to utter) at about 20-30 cm from the reference microphone. 
%The speech was acquired with a \textit{NEUMANN TLM 103} condenser microphone, and later digitalized with a \textit{RME octamic II} audio card. To avoid pop effects, a pop filter was placed between the microphone and the speaker.
%An high-pass filter was also used to mitigate the ``proximity effect'', which causes an over-emphasis of the low frequencies when the microphone is used at close range. 

\subsection{Impulse Response}  \label{sec:irs}
%The other basic ingredient to generate simulated data is the Impulse Response (IR). 
%MO 24/6
%The Impulse Response (IR) 
The impulse response is the most representative feature
characterizing an acoustic space. In the assumption of linear time-invariant reverberant rooms, IRs provide a complete description of the changes
a sound signal undergoes when traveling from one point in space
to another \cite{kutt}. The impulse response can be either measured in the targeted environment or geometrically inferred by simulations. 

Several techniques have been proposed in the last decade for measuring the IR of an acoustic enclosure, including solutions based on Maximum Length Sequence (MLS) \cite{mls}, Linear Chirps \cite{farina}, or Exponential Sine Sweeps (ESS) \cite{farina}. In \cite{Ravanelli-12}, a comparison between these different methods has been proposed for distant speech recognition purposes, showing that ESS outperforms the other methods, especially when long (1 minute) and high dynamic excitation signals can be emitted in the acoustic environment. This result is due to a better management of the harmonic distortions introduced by the loudspeaker as well as to a more favorable SNR.
That study also revealed that using a professional loudspeaker for exciting the acoustic environment (e.g., a professional Genelec 8030) leads to a much more realistic impulse response measurement, if compared to what obtained with a cheaper loudspeaker. Following these guidelines, an IR measurement campaign has been conducted in the context of the DIRHA project to acoustically characterize a real apartment equipped with a network of microphones. As discussed in \cite{lrec}, about 9000 IRs were estimated.

% Piergiorgio
Synthetic room impulse responses can be generated by the well-known Image-source Method (IM) \cite{image}, based on a geometric model accounting for room size, source and microphone positions, and ideal propagation/reflection paths within the enclosure.
The baseline method only considers attenuation and (approximated) time instants of arrival of reflections, which results in quite unrealistic IRs.
Several improvements have been proposed in order
to achieve IRs with characteristics that better match with those
measured in real environments  \cite{Peterson, Lehmann}. In this work, for instance, a modified version of the standard algorithm allowing us to simulate directive sources is considered. This version has shown to be effective to generate IRs better reflecting real-world conditions.
However, such simplified propagation models, assuming an empty shoebox geometry, cannot reproduce the complex patterns of sound propagation in real rooms, as will be shown in Sec.\ref{sec:comp}.

\begin{table*}[t!]
\centering
\tabcolsep=0.11cm
    \begin{tabular}{ | l | c | c | c | c | c | c |}
    \hline
    \multirow{2}{*}{\backslashbox{\em{A.M.}}{\em{Data Type}}} & \multicolumn{2}{ | c | }{\textit{Single Distant Microphone}} & \multicolumn{2}{ | c | }{\textit{Delay-and-Sum Beamforming}} & \multicolumn{2}{ | c | }{\textit{Oracle Microphone Selection}} \\  \cline{2-7}
    & Real Data & Sim Data  & Real Data & Sim Data & Real Data & Sim Data \\ \hline    
    Mono & 62.2 & 64.7  & 56.8 & 58.8 & 49.6 & 51.9 \\ \hline
    Tri1 & 39.8 & 41.1  & 33.9 & 34.9 & 28.0 & 29.2 \\ \hline    
    Tri2 & 33.0 & 33.6  & 28.4 & 29.1 & 22.6 & 23.2 \\ \hline    
    Tri3 & 21.5 & 22.3  & 18.0 & 19.1 & 13.6 & 14.9 \\ \hline    
    Tri4 & 19.9 & 21.4  & 17.5 & 17.4 & 12.6 & 13.8 \\ \hline    
    DNN & 12.0 & 13.2  & 10.7 & 11.6 & 7.2 & 7.6 \\ \hline    
    \end{tabular}
\caption{WER(\%) obtained in a distant-talking scenario with real and simulated data across different acoustic models and microphone processing.}
\label{tab:res}
\end{table*}

\section{Experimental Setup}  \label{sec:expsetup}
%The experimental activity conducted in this work will primarily focus on a comparison between the simulated corpus generated as described in Sec and a real dataset recorded in a real apartment.
This section describes the microphone setup, the task, the corpora as well as the speech recognition framework considered in this work. %for the validation of the data simulation approach proposed in Sec. \ref{sec:datasim}. 

\subsection{Multi-microphone Setup}
\begin{figure}[t!]
\centering
\includegraphics[width=0.49\textwidth]{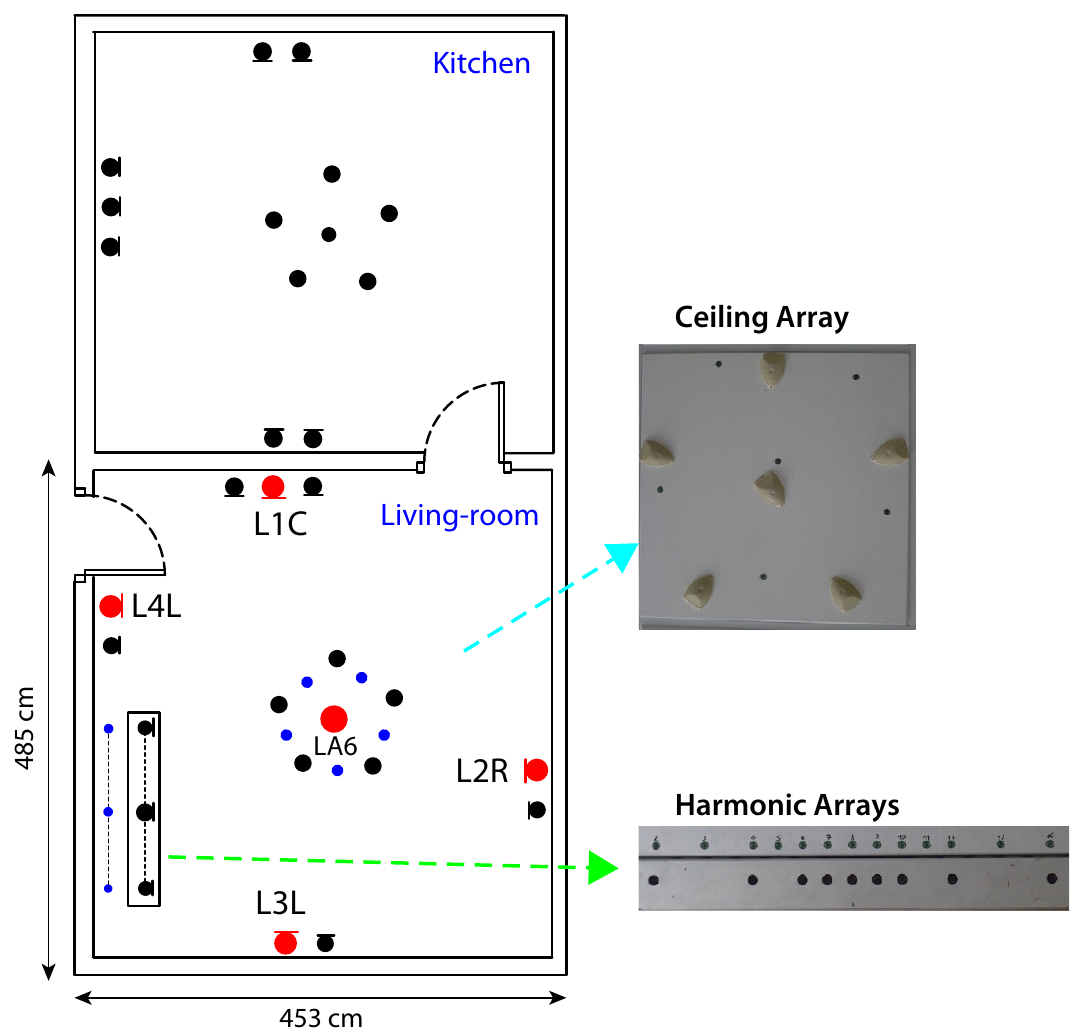}
\caption{An outline of two rooms of the DIRHA apartment considered for this study. Small blue dots represent digital MEMS microphones, red ones refer to the channels considered for the following experimental activity, while black ones represent the other available microphones. The right pictures show the ceiling array and the two linear harmonic arrays installed in the living-room.}
\label{fig:dirhaflat}
\end{figure}

%The reference environment for this study is the living-room  of a real apartment available under the DIRHA project for data collection as well as for the development of prototypes and showcases. 
The apartment used in the DIRHA project is equipped with high-quality omnidirectional microphones (Shure MX391/O), connected to multichannel clocked pre-amp and A/D boards (RME Octamic II), which provide a synchronous sampling at 48 kHz, with 24 bit resolution.
The living-room and the kitchen comprise the largest concentration of sensors and devices. As shown in Fig. \ref{fig:dirhaflat}, the living-room includes three microphone pairs, a microphone triplet, two 6-microphone ceiling arrays (one consisting of MEMS digital microphones), two harmonic arrays (consisting of 15 electret microphones and 15 MEMS digital microphones, respectively). The experiments in this work refer to the use of the five microphones depicted in red in Fig.\ref{fig:dirhaflat}. The reverberation time $T_{60}$ of the considered room is about 0.75 seconds, which indicates that
the acoustic characteristics are quite challenging for DSR studies. 
%Following the methodology described in Sec., a strong effort was devoted to characterize the environment at acoustic level, through different IR estimations considering different positions/orientation of the speaker in reference living-room.

\subsection{Task and corpora}
The task considered in this work is the Wall Street Journal (WSJ-5k), in agreement with the task addressed in the CHiME 3 challenge. While CHiME 3 was pretty focused on robustness against noise, in this work the main source of disturbance is reverberation. 
For testing purposes we employed both real and simulated data, which  are derived from recordings in the DIRHA apartment.
%\footnote{The public distribution of both the real and simulated sentences is being agreed with the Linguistic Data Consortium (LDC).}.
Real data were collected from four native US English speakers (two females and two males) uttering a total of 272 WSJ sentences in different positions of the apartment.
In particular, each subject was positioned in the living-room and read the material from a tablet, standing still or sitting on a chair, in a given position. After a set of 10-12 sentences, she/he was asked to move to a different position and take a different orientation. 
In order to allow a fair comparison between real and simulated data, we asked the same speakers to utter the same sentences in our recording studio, using the acquisition set-up described in Sec.\ref{sec:ctrev}. Moreover, for each position/orientation of the speaker in the real recording, a corresponding IR was measured, allowing us to derive a simulated corpus well-matching with the speaker positions used for the real data. 
The training phase is based on the WSJ0 database (LDC catalog number LDC93S6A), which was contaminated with an impulse response measured in a position different from those used for testing purposes.

\subsection{ASR framework}
The experimental part of this work is based on the Kaldi toolkit \cite{Povey}. The recipe considered for training and testing the DSR system is similar to the s5 recipe proposed in the Kaldi release for WSJ data. In short, the speech recognizer is based on standard MFCCs and acoustic models of increasing complexity. ``\textit{Mono}'' is the simplest system based on 48 context-independent phones of the English language, each modeled by a three state left-to-right HMM (overall using 1000 gaussians). A set of context-dependent models are then derived. In ``\textit{tri1}'' 2.5k tied states with 15k gaussians are trained by exploiting a binary regression tree.``\textit{Tri2}''is an evolution of the standard context-dependent model in which a Linear Discriminant Analysis (LDA) is applied.
In both `\textit{`tri3}'' and ``\textit{tri4}'' models Speaker Adaptive Training (SAT) is also performed. The difference is that  ``\textit{tri4}'' is bootstrapped by the previously computed `\textit{`tri3}'' model.
%The most complex models is ``tri4', which is trained with the Karel's recipe \cite{karel}.
The considered ``\textit{DNN}'', based on the Karel's recipe, is composed of 6 hidden layers of 2048 neurons,  with a context window of 19 consecutive frames (9 before and 9 after the current frame) and an initial learning rate of 0.008. The weights are initialized via RBM pre-training, while the fine tuning is performed with stochastic gradient descent optimizing cross-entropy loss function.

\begin{figure}
\centering
\includegraphics[width=0.43\textwidth]{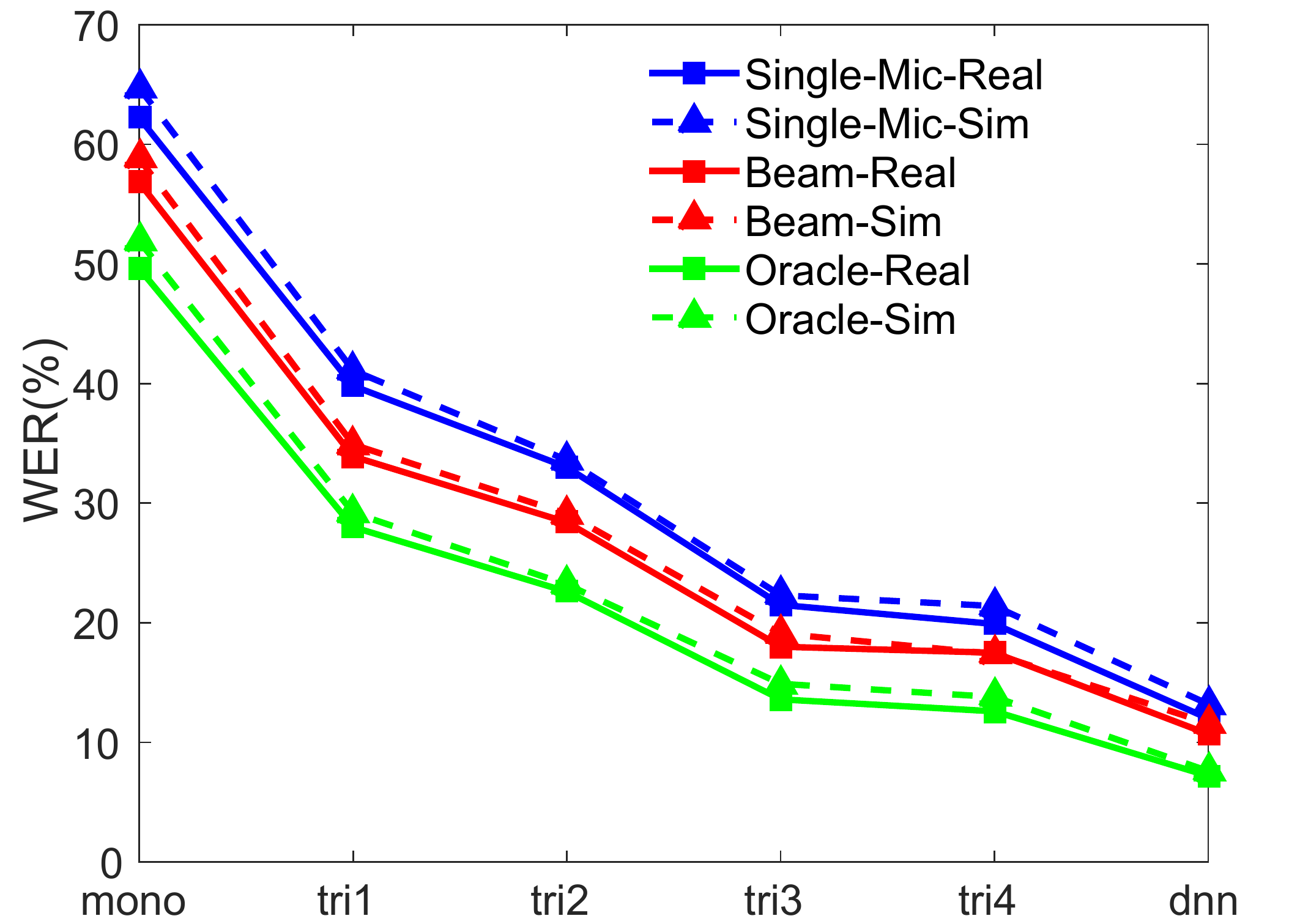}
\caption{Graphical representation of the performance trends reported in Table \ref{tab:res}.}
\label{fig:trend1}
\end{figure}

\section{Results} \label{sec:expres}
This section provides some speech recognition results, with the purpose of validating the proposed data simulation approach. % by comparing real and simulated data over a variety of different experimental conditions.
In the following sub-section, a close-talking baseline is provided, while in subsections \ref{sec:single}, \ref{sec:beam} and \ref{sec:oracle} distant-talking experiments with single microphone, beamforming on the ceiling array, and oracle microphone selection are respectively presented. %Subsection \ref{sec:comp} discusses the results obtained with the image method.

\subsection{Close-talking baseline}
The Word Error Rate (WER\%) obtained by decoding the close-talking WSJ sentences recorded in the recording studio is $3.7\%$ (using DNN models trained with the original clean WSJ data set).  It is worth nothing that, under such favorable acoustic conditions, the DNN model leads to a very accurate sentence transcription.
%MO 24/6
For reference purposes, the average WER with close-talking signals recorded in the DIRHA apartment is about $5\%$.  
%is reported in the first row of Table \ref{tab:ct_res1}.
%As expected, this result highlight that the system performance is significantly improved when passing from a simple monophone-based model to a more competitive DNN baseline. It is however interesting to note that the DNN model leads to a WER of $3.7\%$, which is very close to the error-free condition.% due to both the high quality of the close-talking recordings and to the limited perplexity of the language model.

%Table \ref{tab:ct_res1} 
\begin{figure*}[t!]
\centering
\begin{minipage}{.455\linewidth}
  \includegraphics[width=\linewidth]{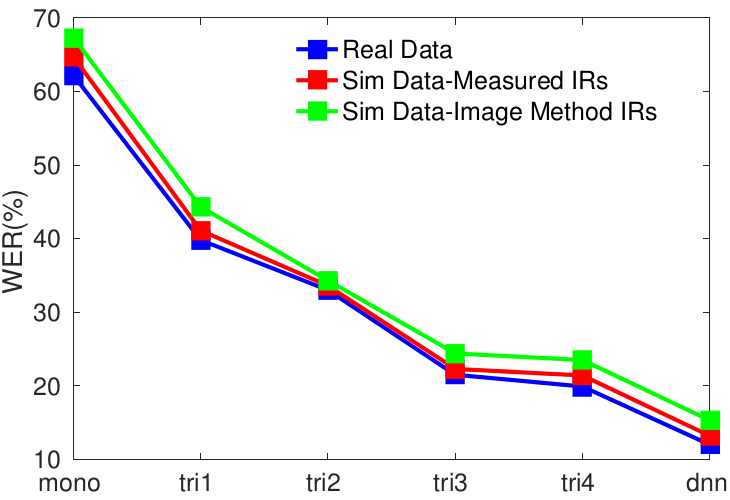}
  \caption{Comparison between real and simulated data with contaminated training performed with a measured IR.}
  \label{img1}
\end{minipage}
\hspace{.05\linewidth}
\begin{minipage}{.46\linewidth}
  \includegraphics[width=\linewidth]{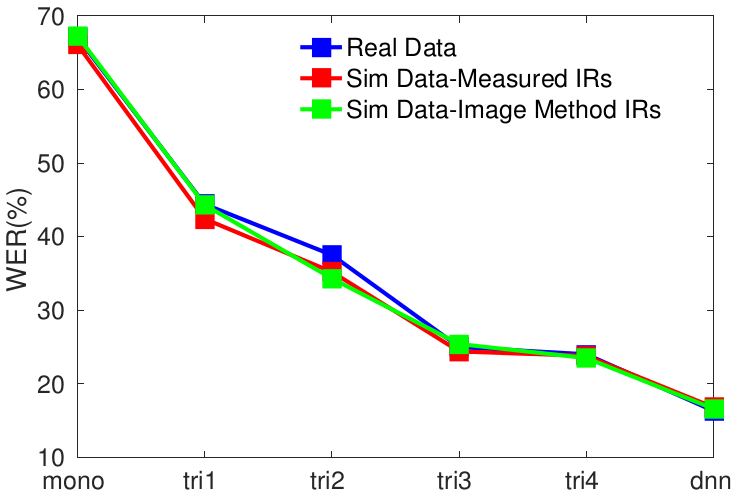}
    \caption{Comparison between real and simulated data with contaminated training performed with an image method IR.}
  \label{img2}
\end{minipage}
\end{figure*}

\subsection{Single distant-microphone performance} \label{sec:single}
The results reported in the first column of Table \ref{tab:res} show the performance obtained when a single distant microphone (i.e., the ``\textit{LA6}'' ceiling microphone depicted in Fig. \ref{fig:dirhaflat}) is considered.
Results clearly highlight that in the case of distant-speech input the ASR performance is dramatically reduced, if compared to a close-talking case. 
The use of robust DNN models trained with contaminated speech material leads, as expected, to a substantial improvement of the WER when compared to other GMM-based systems.
The most interesting result, however, is that a similar performance trend is obtained for both real and simulated data over different acoustic models. This trend can also be appreciated by comparing the continuous (real data) and dashed (sim data) blue lines of Fig. \ref{fig:trend1}. In particular, the average relative WER distance between such data-sets computed over the considered acoustic models is about 6\%. We believe that this is a significant result, especially if one considers that part of this variability can be attributed, despite our best efforts for aligning simulated and real data, to the fact that in the two recording sessions speakers inevitably uttered the same sentence in a different way. % da rivedere questa frase...

\subsection{Delay-and-sum beamforming performance} \label{sec:beam}
The simulation methodology described in Sec. \ref{sec:datasim}, can be extended in a very straightforward way to a multi-microphone scenario. It would be thus of crucial importance to ensure that the similar trend between real and simulated data achieved with a single microphone is preserved even when multi-microphone processing is applied to the data.
Here a standard delay-and-sum beamforming, based on source-microphone delays computed with the GCC-PHAT algorithm \cite{KnappCarter}, is applied to the six microphones of the ceiling array of Fig. \ref{fig:dirhaflat}. 
Table \ref{tab:res} and Fig. \ref{fig:trend1} show that beamforming is helpful in improving the system performance. One can also note that, as hoped, a similar performance trend between the data-sets is reached when applying beamforming. 
For instance, in the case of real data coupled with DNN acoustic models, delay-and-sum beamforming leads to a relative improvement of about 12\% over the single microphone case, which is similar to the improvement of 13\% obtained with the simulated data. %This result may suggest that the proposed data simulation approach is reliable also when a multi-microphone scenario is addressed. 

\subsection{Oracle microphone selection performance} \label{sec:oracle}
To further confirm the result achieved in the previous sections, an oracle microphone selection is applied to both real and simulated data. An oracle microphone selection is performed by selecting, for each sentence uttered by the speaker, the best WER from the five signals acquired by the red microphones in Fig. \ref{fig:dirhaflat}.  %In this section we study the trend over both real and simulated obtained with an oracle microphone selection, in which for each sentence uttered by the speaker the best microphone (among the red channels depicted in Fig. \ref{fig:dirhaflat}) is selected.
Table \ref{tab:res} and Fig. \ref{fig:trend1}, confirm that  the consistency between real and simulated data is largely preserved. The experimental results also show that an optimal microphone selection would be particularly helpful for improving the DSR performance. A proper channel selection has a great potential even when compared with a microphone combination based on delay-and-sum beamforming. For instance, in the case of real data with DNN acoustic models, a WER of 7.2$\%$ is obtained with an oracle channel selection against a WER of 10.7\% achieved with beamforming. 
%24/6
%This experiment, however, only defines an upper bound of the DSR performance, since current microphone selection techniques are still far from this ideal case.

%It is also interesting to note that, assuming a perfect criteria for the selection of the microphone the obtained performance is not so far from that obtained in the close-talking case (WER=3.7\%). . %For instance, in this case a relative improvement of 48\% and $52\%$ is obtained from real-data and simulated data respectively.
%, against a relative improvement of $52\%$ obtained for the simulated corpus.
 
\subsection{Measured vs Geometric Modeling of IRs} \label{sec:comp} % parte da riscrivere meglio
In this section we compare the simulations based on measured IRs, so far considered, with simulations derived by image method-based IRs. For the latter case, the geometry of the targeted living room, the spatial coordinates of microphones and speakers, as well as the reverberation time  $T_{60}$ of 0.75s are imposed to the IM algorithm. As outlined in Sec. \ref{sec:irs}, a certain source spatial directivity similar to that exhibited by a real speaker, is considered. %Moreover, a modified version of the standard Image Method allowing us to simulate directive sources is considered in order to increase the realism of the generated IRs.
Fig. \ref{img1} and Fig. \ref{img2} show the performance observed using two different training strategies. In particular, Fig. \ref{img1} reports the trend obtained when the training set is contaminated with an impulse response measured in the target environment, while Fig. \ref{img2} presents the results obtained when using an image method-based IR. Results confirm that, in both matching and mismatching conditions, simulated data obtained with measured IRs exhibit a trend very similar to that observed with real data. 
%MO 24/6
For instance, in the case of DNN, performance with Real, Sim-Measured IRs, and Image Method, are 12.0\%, 13.2\%, and 15.3\%, respectively.
On the other hand, despite our best efforts for increasing the realism of image method-based IRs, the performance with such
simulation approach is still unsatisfactory.
%data simulation approach is still unsatisfactory.
%worse than that obtained with real data. 
%MO 24/6
In particular, in the case of DNN the relative performance loss using image-method based IRs, instead of measured IRs, for contaminated training is 36\% (i.e., from 12\% to 16.3\%).

%MO 24/6

%Even though deeper analysis on the realism of image method-IRs for DSR experiments should be done as future work, this preliminary study may suggest that simulations based on such methodology 
%represent 
%a drastic simplification of the complex acoustic environment and thus such approach suffers from a lack of realism. 
%is still lacking of realism and efficacy.

% \begin{figure}
% \begin{floatrow}
%  \ffigbox{
% \centering
% \includegraphics[width=0.43\textwidth]{IS4.pdf}
% \caption{Comparison between real and simulated data over different acoustic models.}
% \label{fig:dataset1}}
% \end{floatrow}
% \end{figure}

% \begin{figure}
% \centering
% \includegraphics[width=0.43\textwidth]{IS4.pdf}
% \caption{Comparison between real and simulated data over different acoustic models.}
% \label{fig:dataset1}
% \end{figure}

%         \begin{figure}[!h]
%            \begin{floatrow}
%              \ffigbox{\includegraphics[scale = 0.8]{casxinf1}}{\caption{Case  $ x  < 1 $}\label{case1}}
%              \ffigbox{\includegraphics[scale = 0.8]{zoom}}{\caption{A zoom}\label{zoom}}
%            \end{floatrow}
%         \end{figure}

%\section{Discussion}
   
\section{Conclusion} \label{sec:conc}
In this paper we discussed our best practices to generate realistic multi-microphone data 
%MO 24/6
for training and testing distant-speech recognition systems.
Our approach has been validated by comparing real data with simulated data obtained by convolving close-talking dry speech sequences with impulse responses measured in a domestic environment. The experimental results show that a very similar performance trend can be obtained between real and simulated data over different experimental conditions, involving different acoustic models and multi-microphone processing techniques. This study also revealed that data simulation based on IRs measured in the targeted environment ensures much better results than those obtained with an IR simulator based on Image method. However, in the perspective of a real application,  measuring every time the IRs can be unpractical. The results reported in this paper are thus just a starting point towards a future work, which will study  more in depth how the gap between measured and synthetic IRs can be reduced. An ideal solution would be to automatically analyze the recorded speech and to drive an unsupervised adaptation of initial IRs possibly generated by simulation. 

%for instance involving unsupervised solutions able to automatically analyze the recorded speech to update and improve an initial IM-based IRs.

%\newpage
  \eightpt

\bibliographystyle{IEEEtran}
\bibliography{mybib}

\begin{thebibliography}{10}
\providecommand{\url}[1]{#1}
\csname url@samestyle\endcsname
\providecommand{\newblock}{\relax}
\providecommand{\bibinfo}[2]{#2}
\providecommand{\BIBentrySTDinterwordspacing}{\spaceskip=0pt\relax}
\providecommand{\BIBentryALTinterwordstretchfactor}{4}
\providecommand{\BIBentryALTinterwordspacing}{\spaceskip=\fontdimen2\font plus
\BIBentryALTinterwordstretchfactor\fontdimen3\font minus
  \fontdimen4\font\relax}
\providecommand{\BIBforeignlanguage}[2]{{%
\expandafter\ifx\csname l@#1\endcsname\relax
\typeout{** WARNING: IEEEtran.bst: No hyphenation pattern has been}%
\typeout{** loaded for the language `#1'. Using the pattern for}%
\typeout{** the default language instead.}%
\else
\language=\csname l@#1\endcsname
\fi
#2}}
\providecommand{\BIBdecl}{\relax}
\BIBdecl

\bibitem{BrandWard}
M.~Brandstein and D.~Ward, \emph{Microphone arrays}.\hskip 1em plus 0.5em minus
  0.4em\relax Springer, Berlin, 2000.

\bibitem{beam}
W.~Kellermann, \emph{Beamforming for Speech and Audio Signals}.\hskip 1em plus
  0.5em minus 0.4em\relax in HandBook of Signal Processing in Acoustics,
  Springer, 2008.

\bibitem{nadeu}
M.~Wolf and C.~Nadeu, ``Channel selection measures for multi-microphone speech
  recognition,'' \emph{Speech Communication}, vol.~57, pp. 170--180, Feb. 2014.

\bibitem{bss}
S.~Makino, T.~Lee, and H.~Sawada, \emph{Blind Speech Separation}.\hskip 1em
  plus 0.5em minus 0.4em\relax Springer, 2010.

\bibitem{derev}
P.~A. Naylor and N.~D. Gaubitch, \emph{Speech Dereverberation.}\hskip 1em plus
  0.5em minus 0.4em\relax Springer, 2010.

\bibitem{DeMori98}
R.~DeMori, \emph{Spoken Dialogues with Computers}.\hskip 1em plus 0.5em minus
  0.4em\relax London: Academic Press, 1998, chapter 2.

\bibitem{aed1}
A.~Temko, C.~Nadeu, D.~Macho, R.~Malkin, C.~Zieger, and M.~Omologo, ``Acoustic
  event detection and classification,'' in \emph{Computers in the Human
  Interaction Loop}.\hskip 1em plus 0.5em minus 0.4em\relax Springer London,
  2009, pp. 61--73.

\bibitem{pawel2}
P.~Swietojanski, A.~Ghoshal, and S.~Renals, ``Hybrid acoustic models for
  distant and multichannel large vocabulary speech recognition,'' in
  \emph{Proc. of ASRU 2013}, pp. 285--290.

\bibitem{hain}
Y.~Liu, P.~Zhang, and T.~Hain, ``Using neural network front-ends on far field
  multiple microphones based speech recognition,'' in \emph{Proc. of ICASSP
  2014}, pp. 5542--5546.

\bibitem{dnn_rev}
F.~Weninger, S.~Watanabe, J.~{Le Roux}, J.~Hershey, Y.~Tachioka, J.~Geiger,
  B.~Schuller, and G.~Rigoll, ``{The MERL/MELCO/TUM System for the REVERB
  Challenge Using Deep Recurrent Neural Network Feature Enhancement},'' in
  \emph{IEEE REVERB Workshop}, 2014.

\bibitem{dnn_rev2}
S.~S. Masato~Mimura and T.~Kawahara, ``Reverberant speech recognition combining
  deep neural networks and deep autoencoders,'' in \emph{IEEE REVERB Workshop},
  2014.

\bibitem{dnn3}
A.~Schwarz, C.~Huemmer, R.~Maas, and W.~Kellermann, ``{Spatial Diffuseness
  Features for DNN-Based Speech Recognition in Noisy and Reverberant
  Environments},'' in \emph{Proc. of ICASSP 2015}.

\bibitem{adverse}
E.~H{\"a}nsler and G.~Schmidt, \emph{Speech and Audio Processing in Adverse
  Environments}.\hskip 1em plus 0.5em minus 0.4em\relax Springer, 2008.

\bibitem{ravanelli15}
M.~Ravanelli and M.~Omologo, ``{Contaminated speech training methods for robust
  DNN-HMM distant speech recognition},'' in \emph{Proc. of INTERSPEECH 2015},
  pp. 756--760.

\bibitem{matassoni}
M.~Matassoni, M.~Omologo, D.~Giuliani, and P.~Svaizer, ``{Hidden Markov model
  training with contaminated speech material for distant-talking speech
  recognition.}'' \emph{{Computer Speech \& Language}}, vol.~16, no.~2, pp.
  205--223, 2002.

\bibitem{rav_in14}
M.~Ravanelli and M.~Omologo, ``{On the selection of the impulse responses for
  distant-speech recognition based on contaminated speech training},'' in
  \emph{Proc. of INTERSPEECH 2014}, pp. 1028--1032.

\bibitem{mstyke}
A.~Sehr, C.~Hofmann, R.~Maas, and W.~Kellermann, ``{Multi-style training of
  HMMS with stereo data for reverberation-robust speech recognition},'' in
  \emph{Proc. of HSCMA 2011}, pp. 196--200.

\bibitem{cont2}
L.~Couvreur, C.~Couvreur, and C.~Ris, ``{A corpus-based approach for robust ASR
  in reverberant environments.}'' in \emph{Proc. of INTERSPEECH 2000}, pp.
  397--400.

\bibitem{cont3}
T.~Haderlein, E.~N{\"o}th, W.~Herbordt, W.~Kellermann, and H.~Niemann, ``{Using
  Artificially Reverberated Training Data in Distant-Talking ASR.}'' ser.
  Lecture Notes in Computer Science, vol. 3658.\hskip 1em plus 0.5em minus
  0.4em\relax Springer, 2005, pp. 226--233.

\bibitem{dataaug}
X.~Cui, V.~Goel, and B.~Kingsbury, ``Data augmentation for deep neural network
  acoustic modeling,'' in \emph{Proc. of ICASSP 2014}, pp. 5582--5586.

\bibitem{nakatani}
T.~Yoshioka, N.~Ito, M.~Delcroix, A.~Ogawa, K.~Kinoshita, M.~Fujimoto, C.~Yu,
  W.~J. Fabian, M.~Espi, T.~Higuchi, S.~Araki, and T.~Nakatani, ``{The NTT
  CHiME-3 system: Advances in speech enhancement and recognition for mobile
  multi-microphone devices},'' in \emph{Proc. ASRU 2015}, pp. 436--443.

\bibitem{dataaug2}
S.~R. A.~Ragni, K.~Knill and M.~Gales, ``Data augmentation for low resource
  languages,'' in \emph{Proc. of INTERSPEECH 2014}, pp. 5582--5586.

\bibitem{dataaug3}
T.~Ko, V.~Peddinti, D.~Povey, and S.~Khudanpur, ``Audio augmentation for speech
  recognition,'' in \emph{Proc. of INTERSPEECH 2015}, pp. 3586--3589.

\bibitem{chime}
J.~Barker, E.~Vincent, N.~Ma, H.~Christensen, and P.~Green, ``{The PASCAL CHiME
  speech separation and recognition challenge.}'' \emph{Computer Speech and
  Language}, vol.~27, no.~3, pp. 621--633, 2013.

\bibitem{chime3}
J.~Barker, R.~Marxer, E.~Vincent, and S.~Watanabe, ``{The third CHiME Speech
  Separation and Recognition Challenge: Dataset, task and baselines},'' in
  \emph{Proc. of ASRU 2015}.

\bibitem{revch}
K.~Kinoshita, M.~Delcroix, T.~Yoshioka, T.~Nakatani, E.~Habets, R.~Haeb-Umbach,
  V.~Leutnant, A.~Sehr, W.~Kellermann, R.~Maas, S.~Gannot, and B.~Raj, ``{The
  reverb challenge: A Common Evaluation Framework for Dereverberation and
  Recognition of Reverberant Speech},'' in \emph{Proc. of WASPAA 2013}, pp.
  1--4.

\bibitem{chil_corpus}
D.~Mostefa, N.~Moreau, K.~Choukri, G.~Potamianos, S.~Chu, A.~Tyagi, J.~Casas,
  J.~Turmo, L.~Cristoforetti, F.~Tobia, A.~Pnevmatikakis, V.~Mylonakis,
  F.~Talantzis, S.~Burger, R.~Stiefelhagen, K.~Bernardin, and C.~Rochet, ``{The
  CHIL Audiovisual Corpus for Lecture and Meeting Analysis inside Smart
  Rooms},'' \emph{Language resources and evaluation}, vol.~41, no.~3, pp.
  389--407, 01/2008 2007.

\bibitem{lrec}
L.~Cristoforetti, M.~Ravanelli, M.~Omologo, A.~Sosi, A.~Abad, M.~Hagmueller,
  and P.~Maragos, ``The {DIRHA} simulated corpus,'' in \emph{Proc. of LREC
  2014}, pp. 2629--2634.

\bibitem{dirha_grid}
M.~Matassoni, R.~Astudillo, A.~Katsamanis, and M.~Ravanelli, ``{The DIRHA-GRID
  corpus: baseline and tools for multi-room distant speech recognition using
  distributed microphones},'' in \emph{Proc. of INTERSPEECH 2014}, pp.
  1616--1617.

\bibitem{hscma}
A.~Brutti, M.~Ravanelli, P.~Svaizer, and M.~Omologo, ``A speech event
  detection/localization task for multiroom environments,'' in \emph{Proc. of
  HSCMA 2014}, pp. 157--161.

\bibitem{dirha_icassp}
E.~Zwyssig, M.~Ravanelli, P.~Svaizer, and M.~Omologo, ``A multi-channel corpus
  for distant-speech interaction in presence of known interferences,'' in
  \emph{Proc. of ICASSP 2015}, pp. 4480--4484.

\bibitem{dirha_asru}
M.~Ravanelli, L.~Cristoforetti, R.~Gretter, M.~Pellin, A.~Sosi, and M.~Omologo,
  ``{The DIRHA-ENGLISH corpus and related tasks for distant-speech recognition
  in domestic environments},'' in \emph{Proc. of ASRU 2015}, pp. 275--282.

\bibitem{Ravanelli-12}
M.~Ravanelli, A.~Sosi, P.~Svaizer, and M.~Omologo, ``Impulse response
  estimation for robust speech recognition in a reverberant environment,'' in
  \emph{Proc. of EUSIPCO 2012}.

\bibitem{image}
J.~Allen and D.~Berkley, ``Image method for efficiently simulating small‐room
  acoustics,'' in \emph{J. Acoust. Soc. Am}, 1979, pp. 2425--2428.

\bibitem{kutt}
H.~Kuttruff, \emph{Room acoustic}, 5th~ed.\hskip 1em plus 0.5em minus
  0.4em\relax Spon Press, 2009.

\bibitem{mls}
M.~Schroeder, ``{Diffuse sound reflection by maximum-length sequences},'' in
  \emph{J. Acoust. Soc. Am}, vol. 57(1), 1975, pp. 149--150.

\bibitem{farina}
A.~Farina, ``Simultaneous measurement of impulse response and distortion with a
  swept-sine technique,'' in \emph{Proc. of the 108th AES Convention}, 2000,
  pp. 18--22.

\bibitem{Peterson}
P.~Peterson, ``Simulating the response of multiple microphones to a single
  acoustic source in a reverberant room,'' in \emph{J. Acoust. Soc. Am}, vol.
  80(5), 1986, pp. 1527--1529.

\bibitem{Lehmann}
E.~Lehmann and A.~Johansson, ``Prediction of energy decay in room impulse
  responses simulated with an image-source model,'' in \emph{J. Acoust. Soc.
  Am}, vol. 124(1), 2008, pp. 269--277.

\bibitem{Povey}
D.~Povey~at all, ``{The Kaldi Speech Recognition Toolkit},'' in \emph{Proc. of
  ASRU 2011}.

\bibitem{KnappCarter}
C.~H. Knapp and G.~C. Carter, ``The generalized correlation method for
  estimation of time delay,'' vol.~24, no.~4, pp. 320--327, 1976.

\end{thebibliography}

\end{document}